\documentclass{article}

\usepackage{arxiv}

\usepackage[utf8]{inputenc} 
\usepackage[T1]{fontenc}    
\usepackage{hyperref}       
\usepackage{url}            
\usepackage{booktabs}       
\usepackage{amsfonts}       
\usepackage{nicefrac}       
\usepackage{microtype}      
\usepackage{graphicx}
\usepackage{natbib}
\usepackage{doi}

\usepackage{xcolor}

\title{A New Vertex Connectivity Metric}

\author{ \href{https://orcid.org/0000-0001-9244-8707}{\includegraphics[scale=0.06]{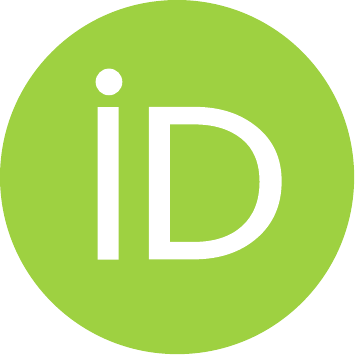}\hspace{1mm}David L.~Rhodes}\\
	US Department of Defense\\
	Fort Meade, MD 20755\\
	\texttt{dlr2dlr2@gmail.com} 
	\And
	\href{https://orcid.org/0000-0002-9084-9459}{\includegraphics[scale=0.06]{orcid.pdf}\hspace{1mm}Breanna N.~Johnson}\thanks{Currently with Applied Physics Laboratory, Johns Hopkins University, Laurel, MD 20723} \\
	US Department of Defense\\
	Fort Meade, MD 20755\\
	\texttt{breanna.johnson1317@gmail.com}
}

\renewcommand{\shorttitle}{\textit{arXiv} Template}

\hypersetup{
	pdftitle={A template for the arxiv style},
	pdfsubject={q-bio.NC, q-bio.QM},
	pdfauthor={David L.~Rhodes, Breanna Johnson},
	pdfkeywords={Networks, Graph, Vertex Connectivity, Algorithm},
}

\usepackage{listings}

\usepackage{algorithm}
\usepackage[noend]{algpseudocode}

\usepackage{booktabs} 

\usepackage{wrapfig}
\newcommand{\eg}{\emph{e.g.}}

\newcommand{\ala}{\emph{ala}}

\newcommand{\proc}[1]{{\sc #1}}

\newcommand{\ns}{\ensuremath{s}}
\newcommand{\na}{\ensuremath{a}}
\newcommand{\nb}{\ensuremath{b}}
\newcommand{\nc}{\ensuremath{c}}
\newcommand{\nr}{\ensuremath{r}}
\newcommand{\nx}{\ensuremath{x}}
\newcommand{\ny}{\ensuremath{y}}
\newcommand{\nz}{\ensuremath{z}}

\newcommand{\Nv}{\ensuremath{v}}
\newcommand{\nt}{\ensuremath{t}}
\newcommand{\vcm}{\texttt{vcm}}

\newcommand{\LS}{\texttt{LS}} 
\newcommand{\IM}{\texttt{IM}} 

\begin{document}
	\maketitle
	
	\begin{abstract}
A new metric for quantifying pairwise vertex connectivity 
in graphs is defined
and an implementation presented.
While general in nature, 
it features a combination of input features well-suited for social networks, 
including applicability to directed or undirected graphs, weighted edges, 
and computes using the impact from all-paths between the vertices.
Moreover, the $O(V+E)$ method is applicable to large graphs. 
Comparisons with other techniques are included.
	\end{abstract}


%
%

\section{Introduction}

A variety of approaches have been developed to answer
the question of interest here, namely: 
\emph{how well connected are a particular pair of vertices?}
This question largely falls into the area of \emph{proximity} measures
which have been previously defined and computed using shortest paths, random walks, diffusion,
maximum flow, similarity, electrical circuits, 
as well as by other means.
Not only are there a multitude of approaches, there are also a varied number
of definitions to quantify the measure as well, often
based on the domain of interest.
We are mostly focused on measures suitable for social network assessment,
but the technique should also be useful in more general network graph settings.

An early measure concerning network connectivity of remotely placed nodes 
or vertices in a graph was developed in
\cite{social:doreian1974}.
The method uses
adjacency matrix multiplications to essentially determine the largest 
flow possible along a single path.
Later work uses multiple shortest paths as a connectivity metric 
\cite{social:ding2008}.
In this effort,
`relationship nodes' are inserted into the natural edges of a social network graph  
to capture common/shared data between the nodes.
%
Vertex pairs are then scored for $k$-vertex-connectivity 
against removal of the added relationship nodes.
This measures the number of `shared-information' connected paths between 
the vertex pair being scored. 
However, a shortcoming with the above methods, is that edge weighting is not included.
Moreover, 
we might not want to restrict a connectivity measure to only 
some number of paths.

The importance of considering 
\emph{all} paths, versus the $k$-shortest, 
single or finite set, is well established in
\cite{social:cohen2012}.
The simplest of the cited measures fall into the category of
route or path accessibility.
The path accessibility metric is the total weight of all paths between
the given source/destination vertices.
\cite{graph:proximity2006}
suggests that the weight of a path could be the \emph{product} of each edge/arc in the
path if weights are adjusted to the $[0 \dots 1]$ range.
When all weights are less than one in a path product formulation, 
more importance is given to shorter paths, 
but such paths are not specifically penalized.
As a variation,
Chebotarev also develops a `Random Forest' accessibility measure, but unfortunately, 
this metric exhibits characteristics that conflict with social network
scoring expectations (see Figs 1 and 2 in \cite{graph:proximity2006}). 
Generally, one
weakness of route and path accessibility is that convergence requires
a fast decrease of proximity value with distances.

Information dispersion, sometimes called rumor dispersion,
in networks is also somewhat related.
In these models, one or more vertices have information that will spread through the network
via neighbor-to-neighbor sharing.
In
\cite{social:chierichetti2010}
a push-pull sharing model is applied as a stepwise algorithm to spread information in networks
and is used, in part, to determine probability bounds on the number of steps needed to complete
information spreading.
Graph \emph{conductance}, a metric developed 
by \cite{social:sinclair1993}, is
identified as a key factor in spreading rate,
although other work
\cite{social:censor2012}
provides specific algorithms to overcome low conductance.
With additional effort, information dispersion between vertices might be developed
into a pairwise connectivity metric, but it would not include edge weighting effects
and would also depend on the spreading model used.

A few other techniques have been developed.
For example, \cite{graph:tong2007}
develops a proximity measure based on an `escape probability'
computed via random walks (see comparison in a later section). 
Rooted page-rank, Katz or eigenvector centrality measures can also be adapted 
for pairwise
connectivity
purposes \cite{social:cohen2012},
again requiring a strong decrease with distance to guarantee convergence.
Faloutsos developed an approach for undirected graphs based on
viewing the graph as an electrical network \cite{graph:faloutsos2004}.
There, extraction of subgraphs is used to provide 
computational efficiency for the solution. 
Connectivity between vertex pairs based on shortest paths has also been studied
\cite{social:xu2004}.
Other `connectivity' metrics are similarity measures \cite{social:newman2018},
which are generally only applicable to nearby vertices.

A survey of proximity measures for social networks can be found in
\cite{social:cohen2012}.
Overall, there does not seem to be an efficient, all-paths method that 
is suitable for possibly very remote vertices in weighted, 
directed
graphs with explicit path weight factors
(that may even inversely emphasize longer paths over shorter ones).

\section{Framing a new metric}\label{framing}

\cite{social:barnes1969} provides an early look at how graph theory can be applied 
to social networks, particularly as related to connectivity while 
\cite{social:martino2006} give a historical perspective on the application of graph 
theory to social network analysis (SNA).
While application of graph theory to SNA has been a subject of much research,
there is still no singular agreement on connectivity metrics.
Therefore, intuitive arguments are made here
to provide a framework for developing
a quantitative metric. 

\begin{figure}[bt]
\centering
\includegraphics[width=4.5in]{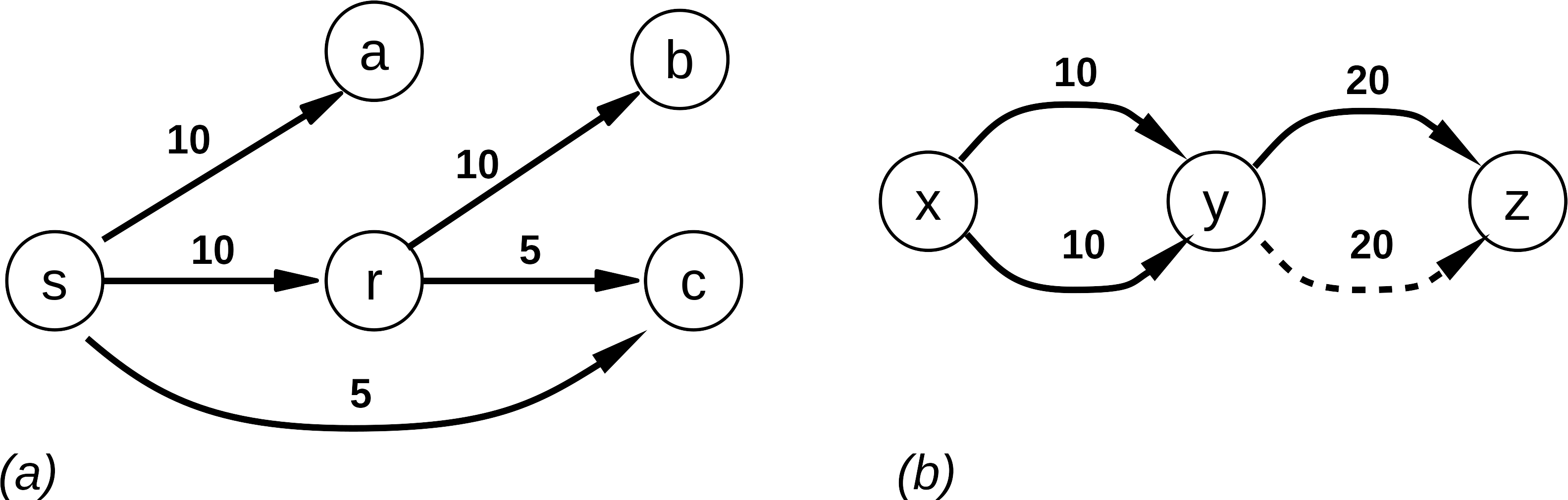}
\caption{Sample graphs}
\label{fig1}
\end{figure}

To illustrate the problem, consider
Figure~\ref{fig1}(a)
which shows a directed graph with weighted edges.
The edge weights are meant to represent a `strength of connection' of the link, 
for example
it might be the number of communication events that have taken place between the entities, 
number of
communication modes, or
the duration of such communications, some combination of both or possibly represent other factors.
The only requirement is that \emph{higher edge-weight values} imply a \emph{stronger} connection;
note that in a sense this is the \emph{opposite} 
of edge weighting with distances where higher values
imply more remoteness.

Consider the connectivity from vertex \ns\ to each vertex \na, \nb, and \nc. 
Intuitively, we would probably say that the \ns\ to \na\ (and \nr) connection is strongest. 
There is a direct connection of `strength' 10.
Would the next strongest connection be between \ns\ and \nb\ or \nc? 
We would argue that the \ns\ to \nc\ connection is the next strongest because there is
a direct connection of 'strength' 5 and an indirect connection of 'strength' 10 
followed by a link of 'strength' 5.
Leaving \ns\ to \nb\ to be the weakest as it \emph{only} has an indirect
connection, even though it is through links of 'strength' 10 and then 10.
Even though there is a total connectivity of 10 from \ns\ to each of \nb\ and \nc, if viewed as a flow,
we would 
likely rate the value of the direct path to \nc\ as more valuable. 

A hallmark of the proposed method is that outgoing edges `divide' the 
attention of a vertex. 
That is the strength of connection present at a vertex is split to its out-neighbors 
in accordance with the numerical
edge weights.
These weight values could be viewed as quantification of strong and weak ties as noted in
\citep{social:granovetter1973}.
As in centrality computations, we will propagate connectivity 
from source to destination
during the calculation. 
A vertex's `score' (connectivity) is propagated
to its neighbors according to outgoing edge weighting.

Next consider Figure~\ref{fig1}(b). The question that arises is
\emph{should the dotted edge add to the connectivity metric from \nx\ to \nz?}
Again intuitively we would say that the connectivity from \ny\ to \nz\ is stronger 
with the dotted edge,
but does (or should) its presence add to the connectivity metric from \nx?
We argue that, in the nominal case where incoming connectivity strengths are summed,
it \emph{should not}.
In this case whether \Nv's attention is divided or not, both routes land at \nt.
Note that this is consistent with a similar argument from \cite{graph:tong2007}.

Suppose there are loops in the graph; how should these impact connectivity
scoring?
We know that
cycles have no impact on the
standard path accessibility value \cite{graph:proximity2006},
but here we argue that each outgoing edge represents a division of attention.
In this case, a loop starting at a vertex should then diminish connectivity along
its other outgoing paths.

These observations imply the following, that the connectivity metric 
should include not only shortest paths between vertices but consider all potential `connectivity flows'.
These flows should consider the number of hops such that
more remote (longer)
connections, even with the same link-by-link connection strength, should not necessarily
be considered equal.
Cycles or loops should have impact, even though a metric like path accessibility does not.

%
%
 
\section{A new vertex-connectivity-metric (VCM)}

Our new \emph{vertex-connectivity-metric} (\vcm)
implements the general characteristics outlined above.
It also is designed to be computationally efficient, and is $O(V+E)$.
As will be seen, the design forms the level graph from the source and 
propagates `connectivity' from a given source vertex along
all paths towards the destination vertex, splitting
connectivity strength and modulating values by level using a given user parameter $\alpha$.  
This asymmetric method is applied to directed, edge-weighted graphs, but
as is common, unweighted graphs can be addressed by making edge weights one and undirected graphs
can be handled by placing edges in both directions.

The \vcm\ algorithm takes as parameters the graph, $G=(V,E)$, 
an `attenuation factor' $\alpha$,
and Boolean user settings:
\LS\ level share,
\IM\ input max,
and the desired source and destination vertices $s,t \in V$ for which the \vcm\ metric is
being computed.
The \LS\ parameter determines if vertices send their connectivity strength along their 
outedges to vertices on the same level,
and \IM\ defines that incoming connectivity strength to the next
level should be the max, not sum, of incoming propagation.
Typical settings might be  
such that the strength of connectivity is summed (\IM\ false), exchange
connectivity strength within a level (\LS\ true) and emphasize shorter paths
($\alpha < 1$).
But one important distinction of the method is allowing 
$\alpha > 1$ to emphasize, rather than de-value, longer paths.

\begin{figure}[bt]
\centering
\includegraphics[height=1.75in]{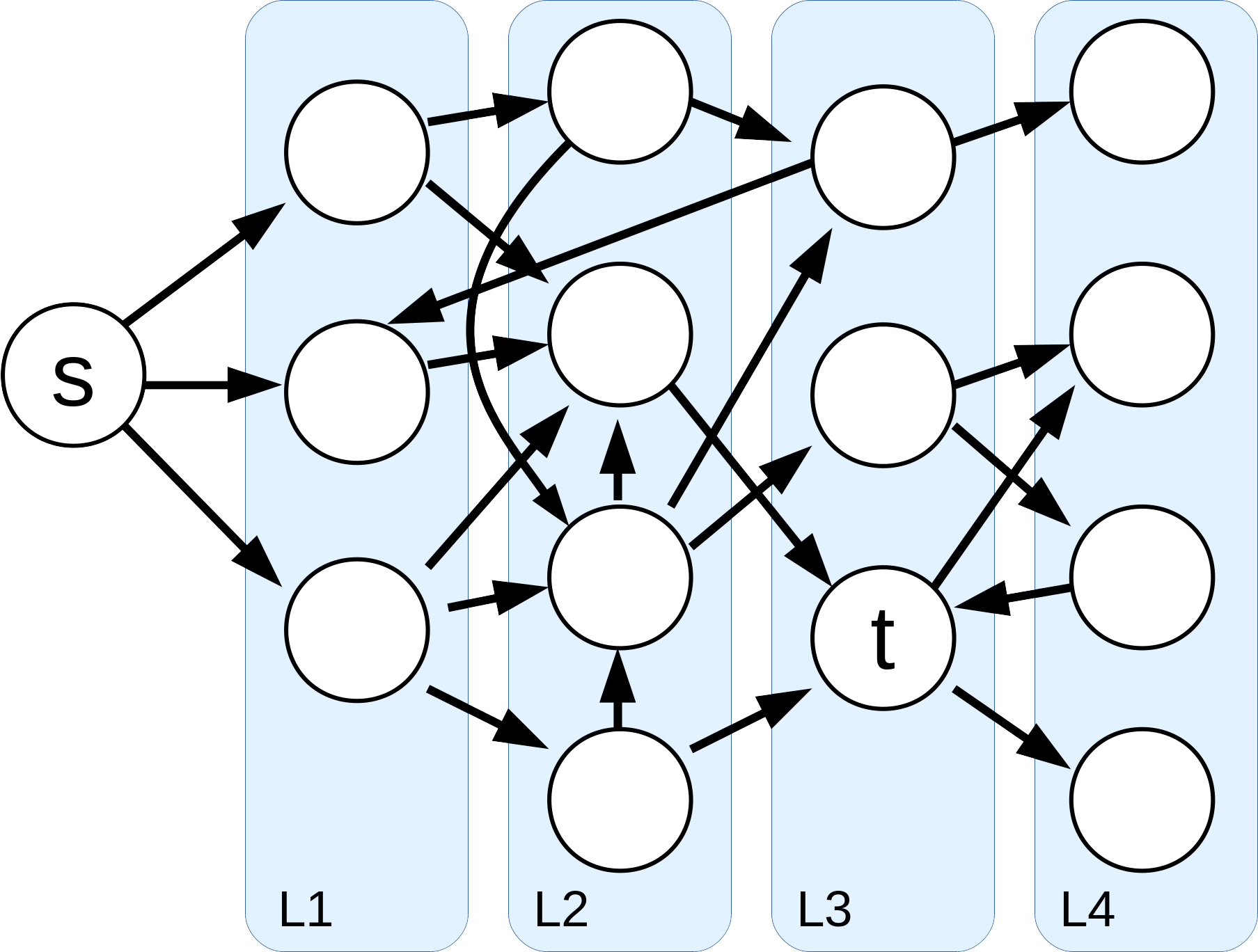}
\hspace{1em}
\includegraphics[height=1.75in]{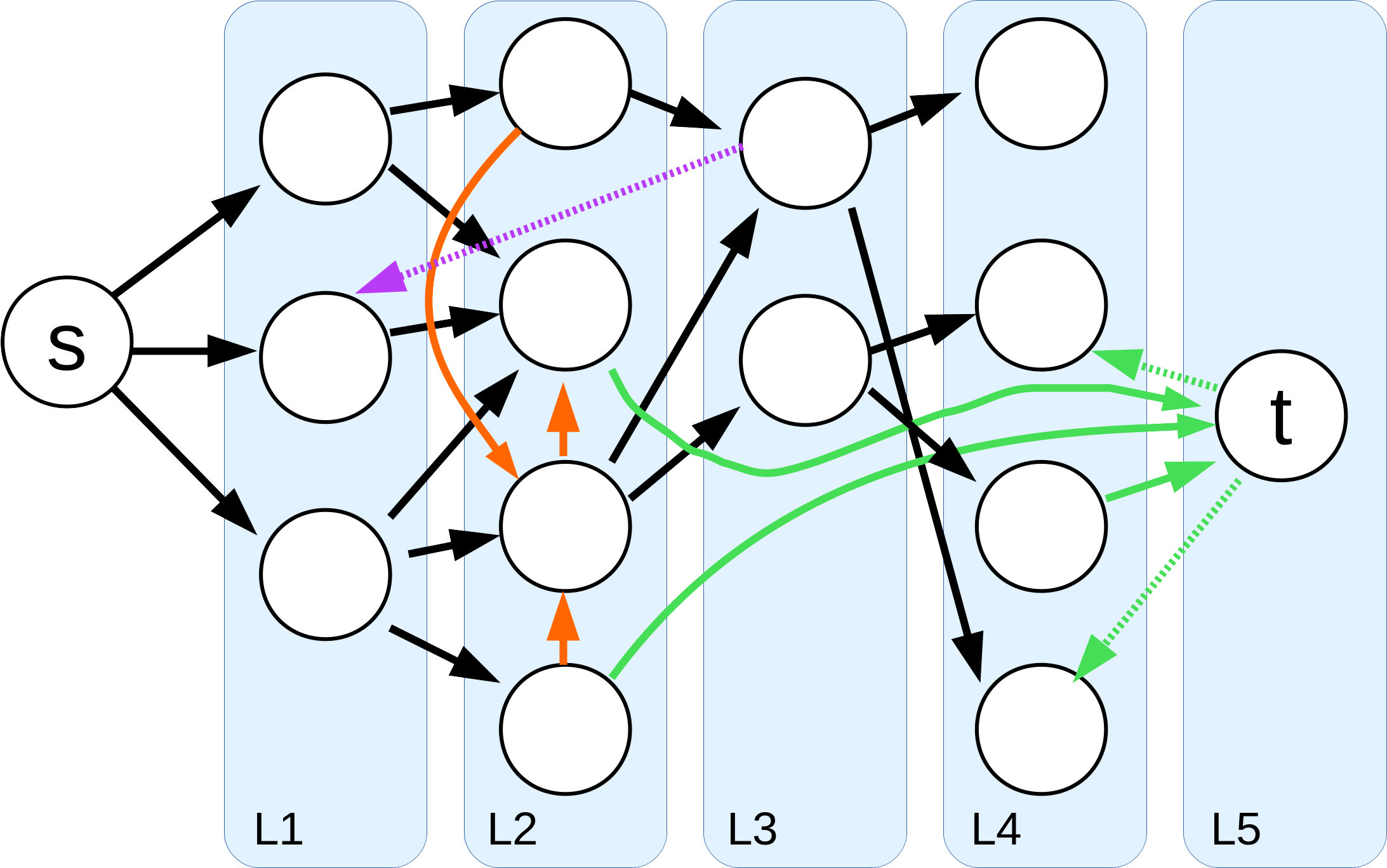}
\caption{Level Graph Formation}
\label{figLevel}
\end{figure}

The routine
creates the level graph from $s$ with the special case that vertex $t$ is `moved' to
a level one higher than all the other vertices.
Figure~\ref{figLevel}
illustrates this process. 
First the input graph is labeled in level graph form, 
as seen on the leftside
of the figure.
The target vertex $t$ is then 
\emph{moved}
to a level beyond the maximum level of all
the other vertices, as seen on the rightside of the figure.
This allows paths that would otherwise be from levels beyond
$t$ to be included in level-by-level connectivity strength propagation to $t$.
For illustrative purposes, the reverse edges (here, L3 to L1 and L5 to L4) 
are shown as dotted, 
the intra-level
edges (in L2) as 
\textcolor{orange}{orange} 
and the edges attached to $t$ in 
\textcolor{green}{green}.

Algorithm~\ref{algvcm}
\begin{algorithm}[tb]
\caption{The \proc{vertex-connectivity-metric} method.}
\label{algvcm}
\begin{algorithmic}[1]
\Procedure{vcm}{$s,t,\alpha$} \Statex \Comment{for simplicity assume $G = (V,E)$ and user settings 
\LS, \IM\ are global}
\If{$s = t$} \Return 1 \Comment max connect to ourself \EndIf 
\State initialize $L[i] \gets -1 \;\; \forall i \in V$ \Comment level of each vertex
\State initialize $W[i] \gets 0.0 \;\; \forall i \in V$ \Comment total weight of all edges from each vertex
\State initialize $M[j] \gets \emptyset$ \Comment set of verts on each level $j$
\State Create level graph using BFS, setting $L, W, M$ in the process
\If{$L[t] = -1$} \Return 0 \Comment $t$ is unreachable from $s$ \EndIf 
\State let $k = $ maximum level and move $t$ to level $k+1$ 
\State $vcm[s] = 1$
\For{$j = 0$ to $k$} \Comment for each level
 \If { \LS\ } \Comment exchange intra-level vcm strength
  \For{$v \in M[j]$}  \Comment for each vertex on level $j$
    \For{edge $v,u$ where $L[u] = L[v]$} \Comment $v$'s neighbors on same level
      \State let $f = w[v,u] / W[v]$ \Comment  weighting over all outedges
      \State $vcm[u] \mathrel{+}= f \times vcm[v]$
    \EndFor
  \EndFor
 \EndIf 

  \State let $a = \alpha^{j}$ \Comment $0^0 = 1$ here
  \For{$v \in M[j]$}  \Comment for each vertex on level $j$, propagate to higher level
    \For{edge $v,u$ where $L[u] > L[v]$}
      \State let $f = a \times w[v,u] / W[v]$ 
      \If{ \IM\ } 
      	\State $vcm[u] = max(f \times vcm[v], vcm[u])$
      \Else
      	\State $vcm[u] \mathrel{+}= f \times vcm[v]$ 
      \EndIf
    \EndFor
  \EndFor
\EndFor

\State \Return $vcm[t]$
\EndProcedure
\end{algorithmic}
\end{algorithm} 
provides an algorithm for computing \vcm.
The \vcm\ metric has a maximum value of 1, and that is returned on line 2 if $s = t$ 
(but see later comment about the maximum exceeding 1 if $\alpha > 1$).
In $O(V)$ time, the level of each vertex is initialized to an `illegal' value of $-1$,
the total out edge weight of each vertex ($W$) is set to zero and the list of vertices on
each level ($M$) is cleared (lines 3-6).

On line 6, a breadth first search is used that sets the level of each vertex (with $L[s] = 0$),
defines the vertex out edge weight total ($W[v] \; \forall v \in V$) and 
sets a list of vertices that are on each level ($M[v]$).
This can be done in $O(E)$ time. On lines 8-9, $t$ is moved to past the highest level and the value of
connectivity for the starting vertex is set to 1.

We then enter a level-by-level loop indexed by $j$ for each level up to $L[t]-1$.
If we are doing level sharing (set by \LS), then VCM scores of vertices 
with links on the same level 
(shown in orange in the figure)
are propagated among each other (lines 11-15). 
This is scaled the same way that scores to the next level are, but is
not subject to the attenuation factor, $\alpha ^j$ defined on line 16.
Next all the connectivity scores on level $j$ are propagated to the next level,
subject to edge-weighting, parameter \IM's settings, and attenuation.
If \IM\ is true, then the propagated connectivity at each vertex 
is set to the maximum of the
incoming transfers, and otherwise it's the sum of these.
The algorithm is written in a straightforward style for clarity.
The routine would be implemented more efficiently by keeping 
separate neighbor lists from each vertex to speed the loops on 
lines 13 and 18.

In this design, there is no propagation along edges to lower levels 
(dotted edges in the figure)
but their weighting
is included in $W$ and diminishes forward (and intra-level) strength propagation.
This does achieve the desired effect outlined in
Section~\ref{framing} 
in that loops should have impact,
but does ignore longer connectivity flows that traverse levels in reverse.

Note that a non-level restricted, Katz-like but edge weighted, 
solution was also considered.
In this case, a `frontier' of vertices, possibly growing to the
size of $V$, would be maintained at each step. At initialization this
would include just $s,1$ (source with strength 1). At each step the frontier's
connectivity would be propagated to their neighbors, including the effects
of $\alpha$, and division \ala\ \IM. After some number of steps,
we would have connectivity measures to all other vertices.
However, there are several issues that we would have to be overcome.
Example, how would the number of steps be defined? There would be oscillation
and non-convergence issues particularly, as in Katz centrality, 
if the $\alpha$ value permitted gain in the system.
Therefore, the simpler level-based approach was devised.

Returning focus to the method,
we can see that this whole routine is $O(V+E)$.
The initialization and BFS routine
to set $W, L, M$ is $O(V+E)$.
With simple neighbor lists, 
the loops from lines 10-23 are $O(E)$ as all graph edges are only processed once each
across all the levels $j$.

After all levels have been so processed, 
$\vcm[t]$ is returned as connectivity strength from $s$ to $t$.
This will be a value from 0 (weakest) to 1 (strongest), with  the
exception that it can exceed 1 when $\alpha > 1$ 
(\eg, there is \emph{amplification} of longer paths). 

We see that this is `single-source/single-destination' routine. 
The need to `move' vertex $t$ to the outmost level in order to capture
longer path contributions prevents using the method as-is for all-destination
purposes. 
But we see that
memoization techniques could be used to save prior
strength propagation values for all levels below some level $k$
while doing the computations for each vertex on level $k$.
That is, moving vertices from level $k$ to the level beyond the maximum
does not change any computed values at levels below $k$.
While this would lead to a more efficient method for an
`single-source/all-destination' version using dynamic programming, 
we leave that for future consideration.

\section{Examples}

For further experimentation, we have produced different outputs
using two different datasets; 
Les-Miserabl\'{e}s character occurrence, 
and Enron email communications.

\begin{figure}[bt]
\centering
\includegraphics[width=6.5in]{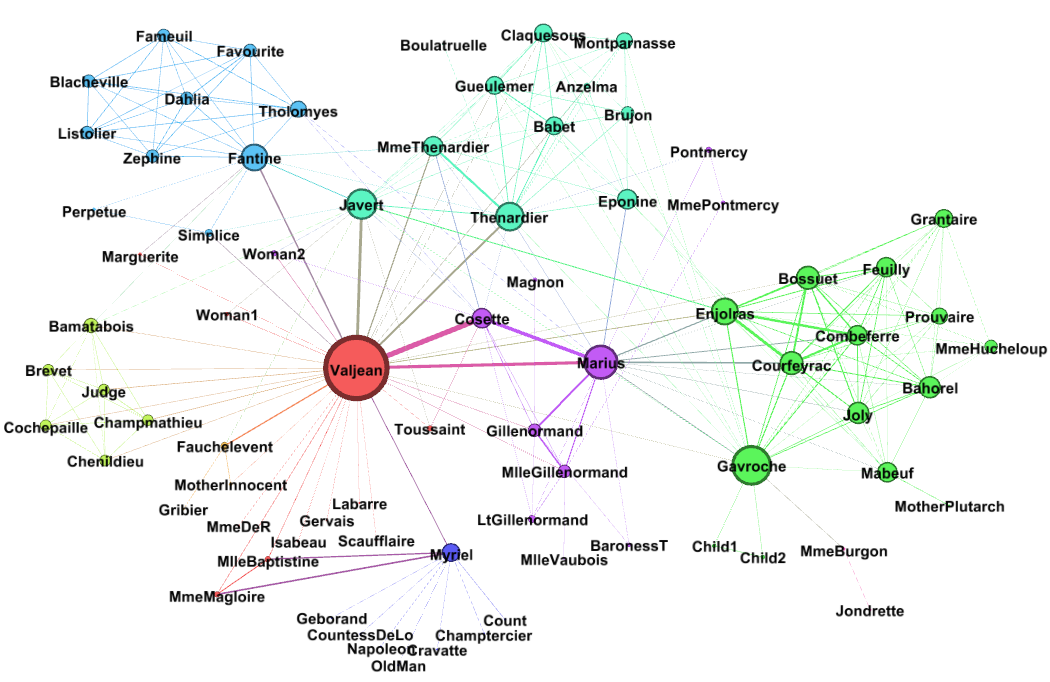}
\caption{Les-Miserabl\'{e}s graph}
\label{figlesmiz}
\end{figure}

Figure~\ref{figlesmiz}
shows the Les-Miserabl\'{e}s undirected network
often used for network analysis demonstration.
Edge weight is the count of the number of times that characters 
appear in the same chapter of the book;
the data can be found at \cite{graph:casos}.
The drawing shows vertices sized by a centrality score, which is not relevant here,
but edge thicknesses are scaled with weight 
giving a visual clue to edge weight values.

We will apply the \vcm\ algorithm from vertex Valjean to all other vertices
for various settings of $\alpha$. 
For this example we will set 
\LS\ true,
\IM\ false.
Table~\ref{tab1}
\begin{table}
\caption{Top ten VCM scoring from Valjean for Les-Miserabl\'{e}s graph}\label{tab1}
\centering
\begin{tabular}{rcccccccccc}
  $\alpha\rightarrow$& 0.0&0.33&0.66&1.0&1.33&1.66&2.0&2.33&2.66&3.0\\
  \hline 
  1 & Cos&Cos&Cos&Cos&Mar&Mar&Mar&Mar&Mar&Mar \\
  2 & Mar&Mar&Mar&Mar&Cos&Cos&Cos&Cos&Enj&Enj \\
  3 & Jav&Jav&Jav&Jav&Jav&The&The&Enj&Bos&Bos \\
  4 & The&The&The&The&The&Jav&Enj&The&Cou&Cou \\
  5 & Fan&Fan&Fan&Fan&Enj&Enj&Jav&Cou&Cos&Com \\
  6 & Fau&Mme&Mme&Mme&Mme&Mme&Cou&Bos&The&The \\
  7 & Mme&Fau&Enj&Enj&Fan&Cou&Bos&Com&Com&Gav \\
  8 & Myr&Myr&Fau&Gil&Cou&Fan&Mme&Jav&Gav&Cos \\
  9 & Enj&Enj&Myr&Fau&Gil&Bos&Com&Gav&Bla&Bla \\
  10 & Cha&Gil&Gil&Myr&Bos&Com&Fan&Mme&Fav&Fav \\
  \end{tabular}
   
\end{table}
shows the top ten vertices (first three characters)
by \vcm\ score for each $\alpha$ value.

Typically, we would set $0 \le \alpha < 1$,
emphasizing stronger connections along shorter paths.
But there is no requirement that $\alpha$ be set less than or equal to 1. By setting to a 
value greater than 1, instead of diminishing the connectivity on longer paths, we would instead
enhance them.
This is useful to finding relatively strong connections to more remote vertices.

At $\alpha = 0$, only direct neighbors get a non-zero score, and their score is
directly in accordance with the outgoing edge weights from $s$ (Valjean). 
In the table, we see that at $\alpha = 0$,
\texttt{Cha} (Champmathieu) is the tenth most connected to Valjean
but it immediately drops off for any other $\alpha$ value.
We see that 
\texttt{Cos} (Cosette) is the highest connectivity vertex until 
$\alpha \ge 1.33$.
From visual inspection of Figure~\ref{figlesmiz}
we see that 
\texttt{Cos} (Cosette) is a direct neighbor of Valjean and also has the 
highest (thicker line) edge weight leading to its top position at lower values
of $\alpha$. 
But at higher values of $\alpha$, 
\texttt{Mar} (Marius) becomes the strongest connection,
this is becuase we are now explicitly emphasizing \emph{longer} paths over shorter
ones, so 
\texttt{Mar} moves into the top spot with multiple strong ties back to Valjean. 
\texttt{Enj} (Enjolras) moves from ninth strongest connection (at $\alpha = 0$) to seventh at
$\alpha = 1$,
and then to second at
$\alpha = 3$.
\texttt{Bos} (Bossuet) is not in the top ten until
$\alpha = 1.33$
and then moves up to third place at
$\alpha = 3$.
Thus, setting $\alpha > 1$ can be seen to help find relatively strongly 
connected vertices that are remote,
while $\alpha < 1$ settings will find strong connections favoring shorter
paths as is typical.

Table~\ref{tab2}
\begin{table}
\caption{Top ten scoring from Valjean for other methods}\label{tab2}
\centering
\begin{tabular}{rccc|cccc}
  &  & &  &\multicolumn{4}{c}{Katz measure} \\
  &comm&tong&mf&$\alpha = 0.33$&$\alpha = 0.66$&$\alpha = 1.00$&$\alpha = 2.00$ \\
  \hline 
  1 & Gav& Cos &Mar &Jav&Jav&Jav&Jav \\
  2 & Enj& Mar &Cos &The&Gav&Gav&Gav \\
  3 & Mar& Jav &The &Gav&The&The&The \\
  4 & Bos& The &Enj &Mar&Mar&Mar&Mar \\
  5 & Cou& Enj &Com &Gue&Enj&Enj&Gue \\
  6 & Jol& Mme &Bos &Bab&Gue&Gue&Bab \\
  7 & Bah& Fan &Cou &Cla&Bab&Bab&Cla \\
  8 & The& Cou &Gav &Mme&Cla&Cla&Mme \\
  9 & Feu& Bos &Jav &Cos&Mme&Mme&Mon \\
  10 & Com& Com &Jol&Enj&Mon&Mon&Epo \\
  \end{tabular}
\end{table}
shows top-ten scoring from other methods.
The metric defined by 
\cite{graph:chen2009} was also considered
but it is not suitable for remote connectivity measures and is more akin to 
a neighbor-based similarity metric.
The column \texttt{comm} is `communicability' from
\cite{social:estrada2008}.
The column \texttt{tong} is the method from
\cite{graph:tong2007}.
The column \texttt{mf} is max-flow, where the connectivity
score is simply the max-flow value.
The remaining ones are the Katz measures as defined in 
\cite{social:cohen2012}. 
A brief discussion of these results follows.

The 
`communicability'  
\cite{social:estrada2008}
result, 
\texttt{comm},
is in the first column of Table~\ref{tab2}.
It is path count based with a value combined from the number of 
shortest paths and the number of longer walks, diminished by length.
It does not account for edge weighting and was originally developed for
community detection applications.
It is the only method that puts 
\texttt{Gav} (Gavroche) as the highest connectivity
(to Valjean) although the katz measures (also unweighted edges) prioritize
that vertex as well.

For the Tong method
(\texttt{tong}), a parameter `c' is used to address numerical issues
related to paths that cannot reach the destination and degree-1 vertices.
The data in the Table uses $c=0.9$ as was also used in the reference.
We see that the scoring is similar to our method,
very closely following it (same first 8 entries) for $\alpha = 1.3$.
This method requires matrix solutions, although a means for finding
all-pairs proximity with one matrix inversion is presented.

Use of maximum-flow as a connectivity measure has been offered.
The scoring for that is in the 
\texttt{mf} column.
One downside is that maximum-flow is flows are contentious 
and therefore it does not include all paths. 
While the value of maximum flow is unique, the flow paths to reach
this value are not. 
In this vein, there doesn't seem to 
be an easy way compose unique sets of constructive paths 
(via Edmonds-Karp or Dinic's algorithms)
without path enumeration (exponential). 
This would make it difficult to apply a path length factor such as
our $\alpha$ in conjunction with flows.

The set of data in Table~\ref{tab2} are the `Katz measures' with 
various $\alpha$ settings.
There are obvious differences versus our method, 
but, as expected, perhaps more commonality with 
\texttt{comm} which is
another method that does not use edge weighting.
When using Katz for centrality, the $\alpha$ term must be smaller than
the reciprocal of the largest eigenvector of the adjacency matrix. 
The metric is based on an infinite sum of walks de-rated on length against
the $\alpha$ term.
For the data presented in
Table~\ref{tab2}, 
the infinite sum was truncated to the diameter of the graph
and hence should capture all non-looping walks 
(and looping ones that are short enough). 
As alluded to in the prior section, if the $\alpha$ term permits
gain, this infinite sum would not converge. 
For the purpose here, the iteration was limited to graph diameter
since we want to explore
various $\alpha$ settings.

Table~\ref{tab3}
\begin{table}
\caption{VCM scoring from Joly}\label{tab3}
\centering
\LS\ true, \IM\ true
\begin{tabular}{rcccccc}
$\alpha\rightarrow$& 0.5&1.0&1.5&2.0&2.5&3.0 \\
\hline 
1&Bab  0.002&Bab  0.003&Fan  0.010&Myr  0.046&Myr  0.175&Myr  0.523\\
2&Bar  0.001&Fan  0.003&Myr  0.008&Fan  0.025&Fan  0.061&Fan  0.184\\
3&Fan  0.000&Myr  0.002&Bab  0.007&Bab  0.015&Bab  0.030&Bab  0.083\\
4&Myr  0.000&Bar  0.001&Bar  0.002&Bar  0.006&Bar  0.011&Bar  0.020\\
\end{tabular}
\\
\vspace{0.5em}
\LS\ true, \IM\ false
\begin{tabular}{rcccccc}
$\alpha\rightarrow$& 0.5&1.0&1.5&2.0&2.5&3.0 \\
\hline 
1&Bab  0.003&Bab  0.015&Fan  0.058&Fan  0.230&Fan  0.745&Fan  2.050\\
2&Fan  0.001&Fan  0.011&Bab  0.048&Myr  0.153&Myr  0.547&Myr  1.584\\
3&Bar  0.001&Myr  0.005&Myr  0.032&Bab  0.125&Bab  0.294&Bab  0.642\\
4&Myr  0.000&Bar  0.002&Bar  0.004&Bar  0.009&Bar  0.015&Bar  0.024\\
\end{tabular}
\\
\vspace{0.5em}
\LS\ false, \IM\ true
\begin{tabular}{rcccccc}
$\alpha\rightarrow$& 0.5&1.0&1.5&2.0&2.5&3.0 \\
\hline 
1&Bab  0.001&Bab  0.001&Fan  0.003&Fan  0.007&Myr  0.021&Myr  0.062\\
2&Bar  0.000&Fan  0.001&Bab  0.002&Myr  0.005&Fan  0.013&Fan  0.030\\
3&Fan  0.000&Bar  0.000&Myr  0.001&Bab  0.003&Bab  0.005&Bab  0.012\\
4&Myr  0.000&Myr  0.000&Bar  0.001&Bar  0.001&Bar  0.003&Bar  0.005\\
\end{tabular}
\\
\vspace{0.5em}
\LS\ false, \IM\ false      
\begin{tabular}{rcccccc}
  $\alpha\rightarrow$& 0.5&1.0&1.5&2.0&2.5&3.0 \\
  \hline 
  1&Bab  0.001&Bab  0.003&Fan  0.012&Fan  0.045&Fan  0.141&Fan  0.380\\
  2&Fan  0.000&Fan  0.002&Bab  0.010&Myr  0.025&Myr  0.089&Myr  0.254\\
  3&Bar  0.000&Myr  0.001&Myr  0.006&Bab  0.024&Bab  0.053&Bab  0.114\\
  4&Myr  0.000&Bar  0.001&Bar  0.001&Bar  0.002&Bar  0.004&Bar  0.006\\
  \end{tabular}
   
\end{table}
illustrates the effect of user settings.
In this case, we are considering the strength of connectivity from
Joly to four others in the graph.
The user controls for 
\LS\ and \IM\ are varied as is the $\alpha$ term.

If we use an $\alpha$ geared to emphasizing remote connections,
\eg\ column $\alpha = 3$,
we see that either 
\texttt{Myr} (Myriel)
or
\texttt{Fan} (Fantine)
are the top choice, depending on the 
\IM\ setting.
If \IM\ is true, then at each level the incoming connectivity
measure to each member is the maximum stemming from the prior level.
If false, then its the sum.
The `effect' of
\IM\ = true
then is to emphasize strongest paths, although all paths are still included.
While for 
\IM\ = false,
all paths contribute at each node as a summation.
So in this case we might say that 
\texttt{Myr} (Myriel)
is better connected via strongly connected paths,
but
\texttt{Fan} (Fantine)
is more connected via a multitude of paths.

%
%

As a final example,
Table~\ref{tab4}
\begin{table}
\caption{Top ten scoring from CEO in Enron data (\LS\ false, \IM\ false)}\label{tab4}
\centering
level 2 vertices
    \vspace{0.2em}
\begin{tabular}{rccccc}
  $\alpha\rightarrow$& 0.5&1.0&1.5&2.0&2.5 \\
  \hline 
  1&susan.m&susan.m&susan.m&veronica.e&veronica.e\\
  2&sgovenar&veronica.e&veronica.e&susan.m&andrew.w\\
  3&miyung.b&sgovenar&cheryl.j&cheryl.j&henry.e\\
  4&tammie.s&miyung.b&tana.j&tana.j&cheryl.j\\
  5&james.s&james.s&miyung.b&sara.s&robert.e\\
  6&kimberly.h&tana.j&sgovenar&andrew.w&sara.d\\
  7&veronica.e&tammie.s&james.s&henry.e&tana.j\\
  8&sharron.w&cheryl.j&sara.s&sara.d&sara.s\\
  9&christi.n&kimberly.h&janette.e&robert.e&susan.m\\
  10&donna.l&janette.e&tammie.s&miyung.b&all.w\\
  \end{tabular}
  \\
  \vspace{0.5em}
  level 3 vertices
    \vspace{0.2em}
  \begin{tabular}{rccccc}
    $\alpha\rightarrow$& 0.5&1.0&1.5&2.0&2.5 \\
    \hline 
    1&sap.h&sap.h&sap.h&pramath\_s&pramath\_s\\
    2&ernest.o&ernest.o&ernest.o&msorrell&msorrell\\
    3&public.r&public.r&msorrell&smu-betas&smu-betas\\
    4&leo.w&leo.w&pramath\_s&hhabicht&hhabicht\\
    5&ken.s&pfranz&public.r&halperin&halperin\\
    6&pfranz&ken.s&leo.w&doyle&doyle\\
    7&traders.e&traders.e&smu-betas&bstephen&bstephen\\
    8&center.e&center.e&jami.h&jami.h&yolanda\\
    9&houston.p&houston.p&pfranz&sap.h&jpainter\\
    10&jeffrey.s&jeffrey.s&ken.s&ernest.o&spikesp\\
    \end{tabular}
    \\
    \vspace{0.5em}
    level 4 vertices 
    \vspace{0.2em}
    
    \begin{tabular}{rccccc}
      $\alpha\rightarrow$& 0.5&1.0&1.5&2.0&2.5 \\
      \hline 
      1&stebbins.j&ken\_lay&ken\_lay&jpfloom&jpfloom\\
      2&ken\_lay&stebbins.j&jpfloom&jmoriarty&jmoriarty\\
      3&canada.d&canada.d&jmoriarty&primlmates&primlmates\\
      4&enron.c&enron.c&canada.d&jgosar&jgosar\\
      5&add...\_ena...&add...\_ena...&enron.c&ken\_lay&20participants\\
      6&glenn.d&glenn.d&add...\_ena...&20participants&christine.m\\
      7&executive.l&executive.l&jgosar&christine.m&dave.d\\
      8&rcunningham&rcunningham&primlmates&dave.d&s20761\\
      9&mmccoy&mmccoy&glenn.d&mgp427&p10621\\
      10&sdaniel&sdaniel&executive.l&s20761&plemme\\
      \end{tabular}
      
\end{table}
provides top ten scoring 
using the Enron email dataset as defined in \cite{social:shetty2004}.
This data has vertices that are email addresses and edges that are emails
between these addresses. 
There are over 2M emails that simplify to
583,550 pairwise (simple) weighted graph edges between
75,153 vertices.
The emails are directed as sender and receiver, however we consider the
graph to be undirected here.
Of course, there can be multiple emails between two parties, and this
count is used as the edge weight.
We could also consider factoring in the email size to compose
a more complex edge weighting scheme, but that is left for future consideration.

For this experiment, we are studying the connectivity from the CEO 
(Jeff Skilling)
to those that are not in direct contact with him.
The first table shows the top ten scores among the set of vertices
that are 2 steps away (\eg, level=2), the next 3 steps and finally 4 steps.
At level=2, there are 15,724 vertices.
At level=3, there are 35,389 vertices and
at level=4, there are 19,080 vertices.\footnote{As Ken Lay was an executive at
Enron, we would expect to find direct communications with Skilling, 
but he uses several email addresses. The one that is coming up on level 4 is
ken\_lay@enron.net}
These tables show the best connected vertices remote from the 
source (Skilling) at given levels 
(also varied against the $\alpha$ term).

%
%

\section{Concluding Remarks}

A new vertex connectivity metric has been developed and demonstrated. 
While geared towards social networks, the metric is general in nature
leveraging only graph structure and weighting.
The method `propagates' connectivity strength values within and to higher
levels in the level-graph image of the input graph.
As loops are avoided, but penalized,
this allows exploration of both gain and loss settings against path length.

A summary of its 
features is:
\begin{itemize}
  \item Leverages edge weighting as a strength of connection value
  \item Applicable to directed or undirected graphs
  \item Level reverse/looping paths are not explicitly included, but
     a penalty is imposed for level-reverse edges 
  \item Path length factors to prioritize either shorter (typical) or longer paths
  \item Efficient, $O(V+E)$ -- does not require matrix solution/inversion.
\end{itemize}

%
%


\bibliographystyle{unsrtnat}
\bibliography{./bibtex/graph,./bibtex/social,./bibtex/other}

\begin{thebibliography}{18}
\providecommand{\natexlab}[1]{#1}
\providecommand{\url}[1]{\texttt{#1}}
\expandafter\ifx\csname urlstyle\endcsname\relax
  \providecommand{\doi}[1]{doi: #1}\else
  \providecommand{\doi}{doi: \begingroup \urlstyle{rm}\Url}\fi

\bibitem[Doreian(1974)]{social:doreian1974}
Patrick Doreian.
\newblock On the connectivity of social networks.
\newblock \emph{J. Mathematical Sociology}, 3:\penalty0 245--258, 1974.

\bibitem[Ding and Dixon(2008)]{social:ding2008}
Li~Ding and Brandon Dixon.
\newblock Using an edge-dual graph and $k$-connectivity to identify strong
  connections in social networks.
\newblock In \emph{{Proc. 46\textsuperscript{th} ACM Southeast Regional Conf.
  }}, pages 475--480, Auburn, AL, March 2008.

\bibitem[Cohen et~al.(2012)Cohen, Kimelfeld, and Koutrika]{social:cohen2012}
Sara Cohen, Benny Kimelfeld, and Georgia Koutrika.
\newblock A survey on proximity measures for social networks.
\newblock \emph{{Lecture Notes in Computer Science (LNCS)}}, pages 191--206,
  January 2012.

\bibitem[Chebotarev and Shamis(1998)]{graph:proximity2006}
P.~Yu Chebotarev and E.V. Shamis.
\newblock On proximity measures for graph vertices.
\newblock \emph{Automation and Remote Control}, 59\penalty0 (10):\penalty0
  1443--59, 1998.

\bibitem[Chierichetti et~al.(2010)Chierichetti, Lattanzi, and
  Panconesi]{social:chierichetti2010}
Flavio Chierichetti, Silvio Lattanzi, and Alessandro Panconesi.
\newblock Rumour spreading and graph conductance.
\newblock In \emph{{Proc. 21\textsuperscript{st} ACM-SIAM Symp. Discrete
  Algorithms (SODA'10)}}, pages 1657--1663, Philadelphia, PA, January 2010.

\bibitem[Sinclair(1993)]{social:sinclair1993}
Allstair Sinclair.
\newblock \emph{Algorithms for Random Generation and Counting: a Markov Chain
  Approach}.
\newblock Birkhauser-Verlag, Basel, Switzerland, January 1993.

\bibitem[Censor-Hillel and Shachnai(2012)]{social:censor2012}
Keren Censor-Hillel and Hadas Shachnai.
\newblock Fast information spreading in graphs with large weak conductance.
\newblock \emph{{SIAM J. Computing}}, 41\penalty0 (6):\penalty0 1451--1465,
  2012.

\bibitem[Tong et~al.(2007)Tong, Koren, and Faloutsos]{graph:tong2007}
Hanghang Tong, Yehuda Koren, and Christos Faloutsos.
\newblock Fast direction-aware proximity for graph mining.
\newblock In \emph{Proc. 13th ACM Knowledge Discovery and Data (KDD'07)}, Aug
  12--15 2007.
\newblock San Jose, CA.

\bibitem[Faloutsos et~al.(2004)Faloutsos, McCurley, and
  Tomkins]{graph:faloutsos2004}
Christos Faloutsos, Kevin McCurley, and Andrew Tomkins.
\newblock Fast discovery of connection subgraphs.
\newblock In \emph{Proc. 10th ACM Knowledge Discovery and Data (KDD'04)}, pages
  118--127, Aug 2004.
\newblock Seattle, WA.

\bibitem[Xu and Chen(2004)]{social:xu2004}
Jennifer Xu and Hsinchun Chen.
\newblock Fighting organized crime: using shortest path algorithms to identify
  associations in criminal networks.
\newblock \emph{Decision Support Systems}, 38:\penalty0 473--487, 2004.

\bibitem[Newman(2018)]{social:newman2018}
Mark Newman.
\newblock \emph{Networks}.
\newblock Oxford University Press, 2018.

\bibitem[Barnes(1969)]{social:barnes1969}
J.~A. Barnes.
\newblock Graph theory and social networks: A technical comment on
  connectedness and connectivity.
\newblock \emph{{Sociology}}, 3\penalty0 (2):\penalty0 215--232, May 1969.

\bibitem[Martino and Spoto(2006)]{social:martino2006}
Francesco Martino and Andrea Spoto.
\newblock Social network analysis: A brief theoretical review and further
  perspectives in the study of information theory.
\newblock \emph{{PsychNology Journal}}, 4\penalty0 (1):\penalty0 53--86, 2006.

\bibitem[Granovetter(1973)]{social:granovetter1973}
Mark Granovetter.
\newblock The strength of weak ties.
\newblock \emph{{Amer. J. Sociology}}, 78\penalty0 (6):\penalty0 1360--1380,
  May 1973.

\bibitem[Carley(2021)]{graph:casos}
Kathleen Carley.
\newblock Center for computational analysis of social and organizational
  systems ({CASOS}).
\newblock \url{http://casos.cs.cmu.edu}, 2021.

\bibitem[{Chen} and {Safro}(2009)]{graph:chen2009}
Jie {Chen} and Ilya {Safro}.
\newblock {A Measure of the Connection Strength Between Graph Vertices with
  Applications}.
\newblock \emph{arXiv e-prints}, September 2009.
\newblock arXiv:0900.4275.

\bibitem[Estrada and Hatano(2008)]{social:estrada2008}
Ernesto Estrada and N.~Hatano.
\newblock Communicability in complex graphs.
\newblock \emph{{Physical Review E}}, 77\penalty0 (3), March 2008.

\bibitem[Shetty and Adibi(2004)]{social:shetty2004}
Jitesh Shetty and Jafar Adibi.
\newblock The {Enron} email dataset: Database schema and brief statistical
  report.
\newblock Technical report, January 2004.

\end{thebibliography}

\end{document}